%% file: samplepaper.tex
\documentclass[runningheads]{llncs}
\usepackage[T1]{fontenc}
\usepackage{graphicx}

\usepackage{xcolor}
\usepackage{algorithm}
\usepackage[noend]{algpseudocode}
\algnewcommand\algorithmicinput{\textbf{Input:}}
\algnewcommand\algorithmicoutput{\textbf{Output:}}
\algnewcommand\Input{\item[\algorithmicinput]}%
\algnewcommand\Output{\item[\algorithmicoutput]}%
\algnewcommand\algorithmicforeach{\textbf{for each}}
\algdef{S}[FOR]{ForEach}[1]{\algorithmicforeach\ #1\ \algorithmicdo}
\let\oldReturn\Return
\renewcommand{\Return}{\State\oldReturn}
\begin{document}
\title{Automatic Digitization and Orientation of Scanned Mesh Data for Floor Plan and 3D Model Generation}
%
%\titlerunning{Abbreviated paper title}
% If the paper title is too long for the running head, you can set
% an abbreviated paper title here
%
\author{Ritesh Sharma\inst{1 \thanks{Worked on this research work during his internship at Palo Alto Research Center.}}\orcidID{0000-0003-1160-3918} \and 
% index{Sharma, Ritesh}
Eric Bier\inst{2}\orcidID{0009-0004-2265-386X} \and
% index{Bier, Eric}
Lester Nelson\inst{2}\orcidID{0000-0001-8556-7644} \and
% index{Nelson, Lester}
Mahabir Bhandari\inst{3}\orcidID{0000-0003-1951-9876} \and
% index{Bhandari, Mahabir}
Niraj Kunwar\inst{3}\orcidID{0000-0002-6345-7652}}
% index{Kunwar, Niraj}
%
\authorrunning{R. Sharma et al.}
% First names are abbreviated in the running head.
% If there are more than two authors, 'et al.' is used.
%
\institute{University Of California, Merced, USA 
\email{rsharma39@ucmerced.edu} \and
Palo Alto Research Center, USA 
\email{\{bier,lnelson\}@parc.com} \and
Oak Ridge National Laboratory, USA 
\email{\{bhandarims,kunwarn1\}@ornl.gov}}
\maketitle              % typeset the header of the contribution
\begin{abstract}
This paper describes a novel approach for generating accurate floor plans and 3D models of building interiors using scanned mesh data. Unlike previous methods, which begin with a high resolution point cloud from a laser range-finder, our approach begins with triangle mesh data, as from a Microsoft HoloLens. It generates two types of floor plans, a "pen-and-ink" style that preserves details and a drafting-style that reduces clutter. It processes the 3D model for use in applications by aligning it with coordinate axes, annotating important objects, dividing it into stories, and removing the ceiling. 
Its performance is evaluated on commercial and residential buildings, with experiments to assess quality and dimensional accuracy.
% The performance of each step is analyzed on commercial and residential buildings, and experiments are conducted to evaluate the appearance of results when different amounts of transparency and numbers of mesh slices are used.
Our approach demonstrates promising potential for automatic digitization and orientation of scanned mesh data, enabling floor plan and 3D model generation in various applications such as navigation, interior design, furniture placement, facilities management, building construction, and HVAC design.
% Our approach has applications in navigation, interior design, furniture placement, facilities management, building construction, and heating, ventilation, and air conditioning (HVAC) design. In general, our approach appears to be promising for automatic digitization and orientation of scanned mesh data for floor plan and 3D model generation.

\end{abstract}
%
%
%

\input{1-introduction}
\input{2-relwork.tex}
\input{3-Methodology.tex}

\input{4-Experiments.tex}

% \input{4-results.tex}
\input{5-conclusion.tex}

\bibliographystyle{splncs04}
\bibliography{references} 
\end{document}

%% file: 1-introduction.tex
%=======================================
\section{Introduction}\label{sec:intro}
%=======================================
Floor plans are useful for many applications including navigating in building interiors; remodeling; efficient placement of furniture; placement of pipes; heating, ventilation, and air conditioning (HVAC) design; and preparing an emergency evacuation plan. Depending on the application, different kinds of floor plan are appropriate. For remodeling building interiors or designing HVAC systems, users may prefer a drafting-style floor plan that focuses on planar walls and removes furniture and other clutter. For furniture placement, navigation, or evacuation planning, users may prefer a more detailed floor plan that shows the positions of furniture, cabinets, counter tops, etc. In either case, producing a floor plan can be time consuming, requiring expert skills, such as measuring distances and angles or entering data into a CAD program. Furthermore, it may need to be done more than once because a building changes when walls and furniture are moved, added, or removed. So it is valuable to be able to generate floor plans automatically with little or no training.

To generate floor plans, it helps to begin with accurate data that can be collected automatically. Laser range finders, smartphones, tablets, and augmented reality (AR) headsets are some of the devices that have made it easier to collect high-resolution building data in the form of RGBD images, point clouds, and triangle meshes. In this paper, we describe a method for generating drafting-style and pen-and-ink-style floor plans by leveraging incomplete and imperfect triangle mesh data. This approach efficiently generates both types of floor plans accurately, supporting a wide range of applications.

\textbf{Main Contribution}
We describe a new method for generating accurate floor plans using poorly captured triangle mesh data from the Microsoft HoloLens 2. The main contributions are:
\vspace{-0.15cm}
\begin{itemize} 
\item A modified Density-Based Spatial Clustering of Applications with Noise (DBSCAN) algorithm, using blocks to capture wall height and thickness.
\item An orientation-based clustering method that finds walls at arbitrary angles.
\item The use of k-means clustering to rotate the mesh to the principal axes and to identify the floor and ceiling.
\item Generating two kinds of precise floor plans from incomplete mesh data.
\end{itemize}

%% file: 2-relwork.tex
%=======================================
\section{Related Work}
%=======================================
\vspace{-0.4cm}
Floor plans are crucial for many applications. Software approaches to floor plan creation depend on data availability and data format. Our work builds on previous research on data collection and floor plan computation.
% Floor plans are important elements in many applications, including navigation, furniture placement, remodeling, sensor placements, and designing HVAC applications. The choice of technique for creating a floor plan depends on the availability of data and on the data format. As detailed in this section, we build on related work in data collection and floor plan computations.

% Are we going to say anything about clustering-based methods?

\textbf{Data collection.} Indoor environments can be captured in many formats, including RGBD images, point clouds, and triangle meshes. Zhang et al.~\cite{zhang2013} uses panoramic RGBD images as input and reconstructs geometry using structure grammars, while~\cite{murali2017} uses a 3D scan to extract plane primitives and generates models using heuristics. A single image is used in some deep learning methods~\cite{dasgupta2016,Hsiao2019,ivan2020,Lee2017,Zou2018} to generate cuboid-based layouts for a single room. Detailed semi-constrained floor plan computations for a complete house require processing a 3D scan of the house~\cite{Liu2018}; the complete scan increases accuracy, but increases computing requirements and time. Pintore and Gobbetti~\cite{Pintore2014} proposed a technique to create floor plans and 3D models using an Android device camera, leveraging sensor data and statistical techniques. Chen et al.~\cite{Chen2022} introduced an augmented reality system utilizing the Microsoft Hololens for indoor layout assessment, addressing intuitive evaluation and efficiency challenges. In our approach, we begin with a triangle mesh from a HoloLens 2, using its Spatial Mapping software~\cite{microsoft_spatial_mapping}, which has been surveyed by Weinmann et al.~\cite{weinmann2021efficient}.
% . Weinmann et al.~\cite{weinmann2021efficient} provide a comprehensive survey of the HoloLens' capabilities for 3D mapping and modeling.

\textbf{Floor plan computations}
Early methods~\cite{Adan2010,Budroni2010,Okorn2010,XIONG2013} relied on image processing techniques, such as histograms and plane fitting, to create floor plans from 3D data. While~\cite{Okorn2010} creates a floor plan by detecting vertical planes in a 3D point cloud,~\cite{Budroni2010} uses planar structure extraction to create floor plans. These techniques rely on heuristics and were prone to failure due to noise in the data.

There has been much progress in floor plan computation using graphical models~\cite{Cabral2014,Furukawa2009,Ikehata2015}. Such models~\cite{Gao2014} are also used to recover layouts and floor plans from crowd-sourced images and location data. One interactive tool~\cite{Liu2013} creates desirable floorplans by conforming to design constraints.

Pintore et al.~\cite{Pintore2020} characterizes several available input sources (including the triangle meshes that we use) and output models and discusses the main elements of the reconstruction pipeline.  It also identifies several systems for producing floor plans, including FloorNet~\cite{Liu2018}, and Floor-SP~\cite{Chen_2019}.

Monszpart et al.~\cite{Monszpart2015} introduced an algorithm that exploits the observation that distant walls are generally parallel to identify dominant wall directions using k-means. Our approach also utilizes k-means, but does so to identify walls in all directions, not just the dominant ones.

Cai et al.~\cite{cai2022} uses geometric priors, including point density, indoor area recognition, and normal information, to reconstruct floorplans from raw point clouds.

In contrast to Arikan et al.~\cite{Arikan2013}, which employed a greedy algorithm to find plane normal directions and fit planes to points with help from user interaction, our approach is automatic. It also differs from the work of~\cite{mura2016}, which focuses on removing clutter and partitioning the interior into a 3D cell complex; our method specifically divides the building into separate walls.

Our work is related to ~\cite{Okorn2010} and~\cite{Turner2012}. In~\cite{Okorn2010}, floor plan generation starts with a laser range data point cloud, followed by floor and ceiling detection using a height histogram. The remaining points are projected onto a ground plane, where a density histogram and Hough transform are applied to generate the line segments that form a floor plan.
% ~\cite{Okorn2010} begins with laser range data that produces a point cloud and then uses a histogram of height data to detect the floors and ceilings. After ceiling and floor removal,~\cite{Okorn2010} projects the remaining points onto a 2D ground plane, creates a histogram of point density, and performs a Hough transform to produce line segments in 2D that form a floor plan. 
In projecting to 2D, their method risks losing information that may be useful for creating 3D models or detailed floor plans. Similarly, ~\cite{Turner2012} utilizes a histogram-based approach to detect ceilings and floors. Their method involves identifying taller wall segments to create a 2D histogram, and then employing heuristics based on histogram point density to compute the floor plan.
% Similarly, ~\cite{Turner2012} uses a histogram approach to detect ceilings and floors, but differs in its method for computing the floor plan. Their approach looks for wall segments that are taller than a threshold to compute a 2D histogram and then applies heuristics based on the density of points in the neighborhood of a histogram cell to compute the floor plan. 
Our approach differs from~\cite{Okorn2010} and~\cite{Turner2012} by aligning the mesh with global coordinate axes and not relying on laser data or a point cloud. Working primarily with 3D data throughout the pipeline, it benefits from enhanced information and generates both a 3D model and a floor plan.
% it includes steps to align the mesh with the global coordinate axes and does not require laser data or a point cloud. Starting with a mesh, it works with 3D data for most of the pipeline so that it has more information. It produces both a 3D model and a floor plan.
\vspace{-0.4cm}

%% file: 3-methodology.tex
%=======================================
\section{Methodology}
%=======================================
\vspace{-0.4cm}
We compute floor plans in four main steps (see Algorithm~\ref{alg:OverallApproach}). First, a user captures the interior of the building as a triangle mesh using an augmented reality headset. The mesh is oriented to align with primary axes, and the building is divided into stories. Floors and ceilings are removed, and flat walls are detected if desired. Finally, one of two floor plan styles is generated by slicing and projecting the resulting 3D model. Next we describe these steps in detail.

\textbf{Data collection}
Indoor environments can be captured in various formats using different devices. We use a Microsoft HoloLens 2 headset to capture triangle mesh data, annotating the mesh using voice commands.

\begin{figure}[!ht]
  \includegraphics[width=0.6\columnwidth]{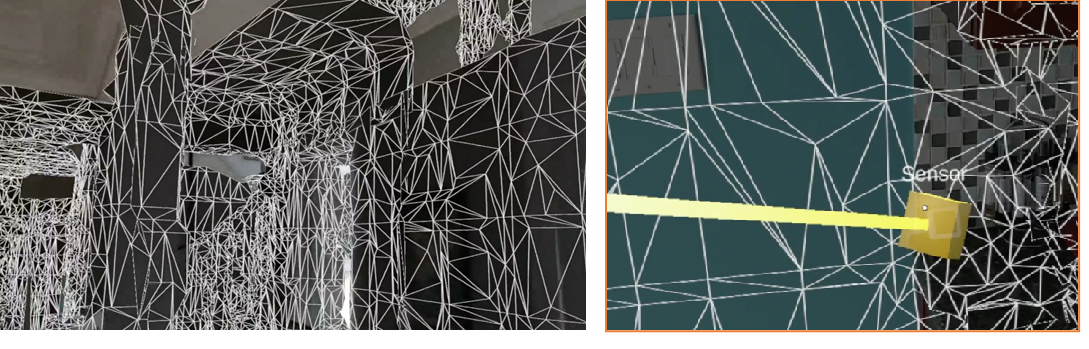}
  \centering
  \vspace{-0.35cm}
  \caption{Left: A triangle mesh is being computed by a HoloLens. Right: The yellow block marks a sensor position. The thick yellow line, known as the \textit{probe}, shows the proposed position and orientation of a sensor or other annotation.}
    \label{fig:hololens2}
\end{figure}

% \vspace{-0.8cm}
\textbf{Capturing the triangle mesh} The HoloLens provides hardware and software to create a 3D representation of the indoor environment using triangles, as shown in Figure~\ref{fig:hololens2}-left. The headset overlays the triangles on the user's view of the building interior. Although the headset captures most of the walls, floors, and ceilings, data may be missing from some regions, as in the figure.

\vspace{-0.3cm}
\begin{algorithm}[!htb]
\footnotesize{
\caption{Methodology}
\begin{algorithmic}[1]
  \State $\mathcal{M} \gets$ \textit{CaptureData}$(\mathcal{E})$;
  \State $\mathcal{PM} \gets$ \textit{MeshOrientation}  $( \mathcal{M} )$; 
   \State $\mathcal{RM} \gets $ \textit{RemoveCeilingsAndFloors} $(\mathcal{PM})$;
    \State $\mathcal{FP} \gets $ \textit{ComputeFloorPlan} $(\mathcal{RM})$;
\Return{$\mathcal{FP}$}.
\end{algorithmic}
\label{alg:OverallApproach}
}
\end{algorithm}

\vspace{-1.1cm}
\vspace{-0.75cm}

\textbf{Annotating the mesh} 
To capture object positions such as sensors, thermostats, windows, and doors, we developed an augmented reality (AR) user interface. This interface uses eye gaze detection and voice commands to enable users to place synthetic objects at desired locations, as shown in Figure~\ref{fig:hololens2}-right where a synthetic sensor object is added to the immersive environment, superimposed on a physical sensor.

% To capture the positions of objects such as sensors, thermostats, windows, and doors, we have designed an augmented reality user interface, as the HoloLens does not automatically capture such objects. Our interactive interface uses eye gaze detection to enable the user to place a dot at a location of interest. While looking at the object, the user gives voice commands indicating the type of object (e.g., sensor) and indicating when the eye gaze point is positioned correctly. The user then instructs the system to add a synthetic object at the specified position. Figure~\ref{fig:hololens2}-right illustrates a synthetic sensor object added to the immersive environment; it is superimposed on a physical sensor.
\vspace{-0.4cm}
\subsection{Mesh orientation}
After the annotated triangle mesh is captured, geometric processing is performed. 
% to prepare it for use. 
Initially, the mesh's orientation is based on the user's starting position and gaze direction. To generate a floor plan, we must determine the floor's position and facing direction. The AR headset provides a rough estimation of gravity direction, but additional computation improves precision.
\vspace{-0.5cm}
% After capturing the triangle mesh with annotations, further processing is necessary to prepare it for use. Initially, the mesh's orientation is based on the user's starting position, i.e., their gaze direction when starting to use the headset. However, to generate an accurate floor plan, we must determine the position and facing direction of the floor. The augmented reality headset can already determine the rough direction of gravity using its built-in sensors, but additional computation is needed to improve its precision.

\subsubsection{Orienting the floor}
To determine the mesh orientation, we tested two methods: (1) compute the shortest edges of the mesh bounding box, and (2) cluster the facing directions of mesh triangles using spherical k-means. 
% we tested two methods for finding the negative y-axis (gravity direction): 
% (1) we compute the bounding box of the mesh and use the direction of its shortest edges, and (2) we cluster the facing directions of the mesh triangles using spherical k-means. 
Method (1) works for buildings with constant altitude and large floor area, but it fails on others, so we mainly use Method (2), described in Algorithm~\ref{alg:SKMMethod}. 

Algorithm~\ref{alg:SKMMethod} applies to a broad range of meshes, including multi-story buildings with vertical dominance.
% but it works for a wider range of meshes, including multi-story building models that are larger vertically than horizontally. 
It uses the surface normal vector of each triangle $\Delta$ in the mesh $\mathcal{M}$ and filters out triangles deviating significantly from the positive y direction, preserving those likely to represent the floor (${\Delta}^{'}$).
% This method starts with the triangle mesh and considers the surface normal vector of each triangle $\Delta$ in the mesh $\mathcal{M}$. We discard any triangles where the facing direction deviates significantly from the positive y direction, retaining only those that are most likely to be part of the floor (${\Delta}^{'}$).

We use a spherical coordinates k-means algorithm with $k = 1$ to find the dominant direction $g_{m}$ of these triangles. We discard triangles that are more than an angle $\phi$ from the dominant direction and repeat the k-means algorithm until $\phi_{min}$ is reached (e.g., start with $\phi =30$ degrees and end with $\phi_{min} =3$). This gives an estimate of the true gravity direction $g_{t}$.

To orient the mesh, we compute the angle $\theta$ between $g_{t}$ and the negative y-axis and determine the rotation axis $\mathcal{Y}$ by taking their cross product. We rotate the mesh by $\theta$ around the $\mathcal{Y}$ axis, ensuring a horizontal floor. Further details on this method for floor orientation are in Algorithm~\ref{alg:SKMMethod}.

\begin{figure}[!ht]
  \includegraphics[width=0.45\columnwidth]{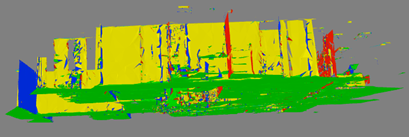}
  \vspace{-0.35cm}
  \centering
  \caption{A model where the floor is not level initially}
  \label{fig:tiltedmodel}
\end{figure}

Figure~\ref{fig:tiltedmodel} shows a model where the floor is not level, but tilts down from near to far and from right to left. After Algorithm~\ref{alg:SKMMethod}, the floor is horizontal.
\vspace{-0.5cm}

\subsubsection{Finding the height of the floor} \label{sec:findheight}
After orienting the mesh to have a horizontal floor, we find the altitude of the floor in the y direction: we take the centroid of each mesh triangle whose facing direction is within a small angle of the positive y axis. We create a histogram of the y coordinates of these centroids, with each bucket representing a vertical range, such as 2 inches. We consider adjacent pairs of buckets and look for the pair with the highest number of points, such as (0, 1), (1, 2), etc. For a single-story building, we search for two large bucket pairs representing the floor (near the bottom) and the ceiling (near the top).

% we compute the axis of rotation $\mathcal{Y}$ by taking the cross product of those vectors. We then rotate the mesh by $\theta$ around the $\mathcal{Y}$ axis. This ensures that the floor is horizontal. More information about the spherical k-means method for orienting the floor can be found in Algorithm~\ref{alg:SKMMethod}.

\vspace{-0.35cm}
\begin{algorithm}[!htb]
\footnotesize{
\caption{Orienting the floor using the Spherical k-means method}
\begin{algorithmic}[1]
    \ForEach {$\Delta \in \mathcal{M} $}
        \If { $N$ of $\Delta$ facing into the room}
            \State ${\Delta}^{'} \gets \Delta$;
        \EndIf
    \EndFor
    \While {$\phi < \phi_{min}$}
        \State ${\mathcal{\phi}, g_{m}}~\gets$~\textit{SphericalKmeans}($\Delta^{'},1$);
             \ForEach {$N$ \textit{of} $\Delta \in \mathcal{M} $}
                \If {\textit{angle}$(N) < \phi$}
                    \State ${\Delta}^{'} \gets \Delta$;
                \EndIf
            \EndFor
    \EndWhile
   \State $\vec{g_{t}}~\gets$ -$\vec{y}$;
  \State $\theta~\gets \arccos(\vec{g_{t}}$ . $\vec{g_{m}})$;
  \State $\mathcal{Y}~\gets \vec{g_{t}}$ x $\vec{g_{m}}$;
  \State $\mathcal{M}~\gets \textit{rotateMesh}(\theta, \mathcal{Y})$;
\end{algorithmic}
\label{alg:SKMMethod}
}
\end{algorithm}

\vspace{-1.0cm}
\vspace{-0.5cm}

% \subsection{Removing Floors and Ceilings}
% To compute a floor plan and view a multilevel building, it's necessary to remove ceilings and floors as desired by the user. The first task is to find the height of the floor and ceiling and divide the building into separate levels. This process also helps to compute the height of the wall and generate 3D models from mesh data.

If the building has sunken floors or raised ceilings, the histogram will show spikes at similar but not identical altitudes. To ensure that we locate true ceilings and floors, we search for a gap of several feet (such as the expected floor-to-ceiling height of a room) between the low and high histogram spikes. The spikes below this gap are probably floors, and those above are probably ceilings.

To generate the floor plan, we choose the highest of the floor levels and the lowest of the ceiling levels as the computed floor and ceiling levels, respectively. Pairing the buckets rather than taking them individually ensures that we do not overlook spikes in the histogram if the mesh triangles are distributed evenly across two adjacent buckets.

\vspace{-0.25cm}
\subsubsection{Rotate mesh and associated annotations}
Our next goal is to align the mesh model's primary wall directions with the axes of Euclidean coordinates.

One optional step is to eliminate mesh triangles whose surface normals are within a small angle from the positive or negative y directions, as these are probably ceiling or floor triangles. This step is not mandatory, but decreases the number of triangles to be processed. Additionally, we eliminate all triangles below the computed floor altitude and all above the computed ceiling altitude.

\begin{figure}[!htb]
  \includegraphics[width=1.0\columnwidth]{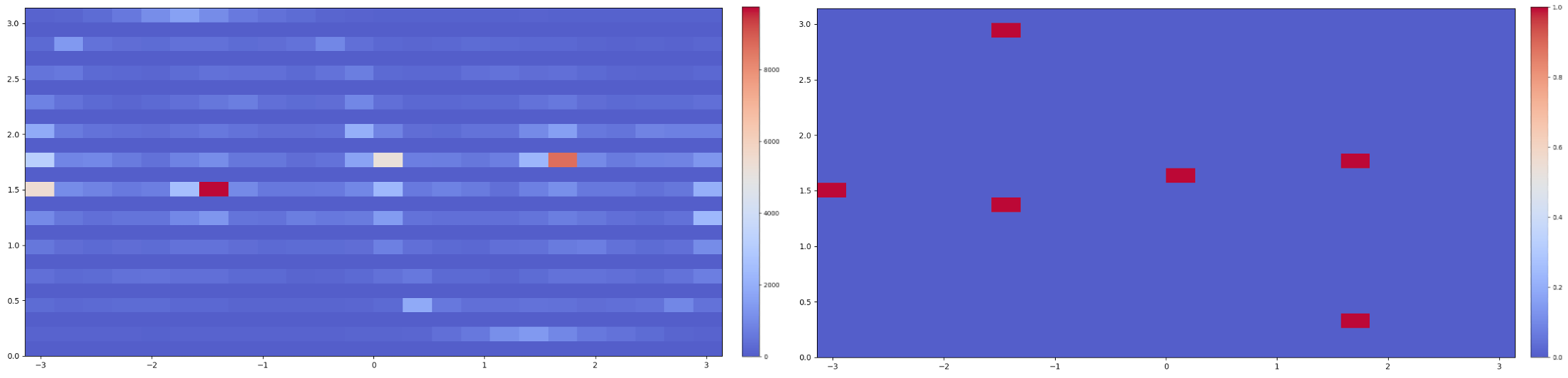}
  \vspace{-0.6cm}
  \centering
  \caption{Left: A heat map of triangle facing directions in spherical coordinates. The horizontal axis is the angle $\theta$ around the y-axis. The vertical axis is the angle $\phi$ from the south pole to the north pole. Warmer colors indicate more triangles per direction bucket. Right: The six cluster centers show the directions of the ceiling (top red dot), the floor (bottom red dot) and the four primary wall directions (dots running left to right at medium height).}
  \label{fig:heatmap1}
\end{figure}

We then examine the surface normals of the remaining triangles. We express each normal in spherical coordinates and use spherical k-means clustering to identify the dominant wall directions. Assuming the building has mainly perpendicular walls, there will be four primary wall directions, so we can set $k = 4$ for k-means clustering. If the model still has floor and ceiling triangles, we can set $k = 6$ to account for the two additional primary directions. Figure~\ref{fig:heatmap1}-left illustrates a heat map of surface normal directions in spherical coordinates from an office building mesh.

\vspace{-0.4cm}
\vspace{-0.75cm}

Figure~\ref{fig:heatmap1}-left contains many light blue rectangles that are far from any cluster center (e.g., far from the buckets that are red, orange, and white). These represent triangles whose facing directions do not line up with any of the primary walls, floors, or ceilings. Such triangles exist for two reasons: (1) Building interiors contain many objects that are not walls, floors, or ceilings, such as furniture, documents, office equipment, artwork, etc. These objects may be placed at any angle. (2) The AR headset generates triangles that bridge across multiple surfaces (e.g., that touch multiple walls) and hence point in an intermediate direction. To compensate, we use a modified version of spherical coordinates k-means clustering that ignores triangle directions that are outliers as follows:

After computing spherical k-means in the usual way, we look for all triangles in each cluster whose facing direction is more than a threshold $\theta_1$ from the cluster center. We discard all such triangles. Then we run k-means again, computing updated cluster centers. Then we discard all triangles that are more than $\theta_2$ from each cluster center where $\theta_2 < \theta_1$. We repeat this process several times until we achieve the desired accuracy. For example, in our current implementation, we use this sequence of angles $\theta_i$ in degrees: [50, 40, 30, 20, 10, 5, 3]. Once our modified k-means algorithm completes, we have 4 (or 6) cluster centers. Figure~\ref{fig:heatmap1}-right shows sample results for $k = 6$.

\vspace{-0.3cm}
\begin{figure}[!ht]
  \includegraphics[width=0.5\columnwidth]{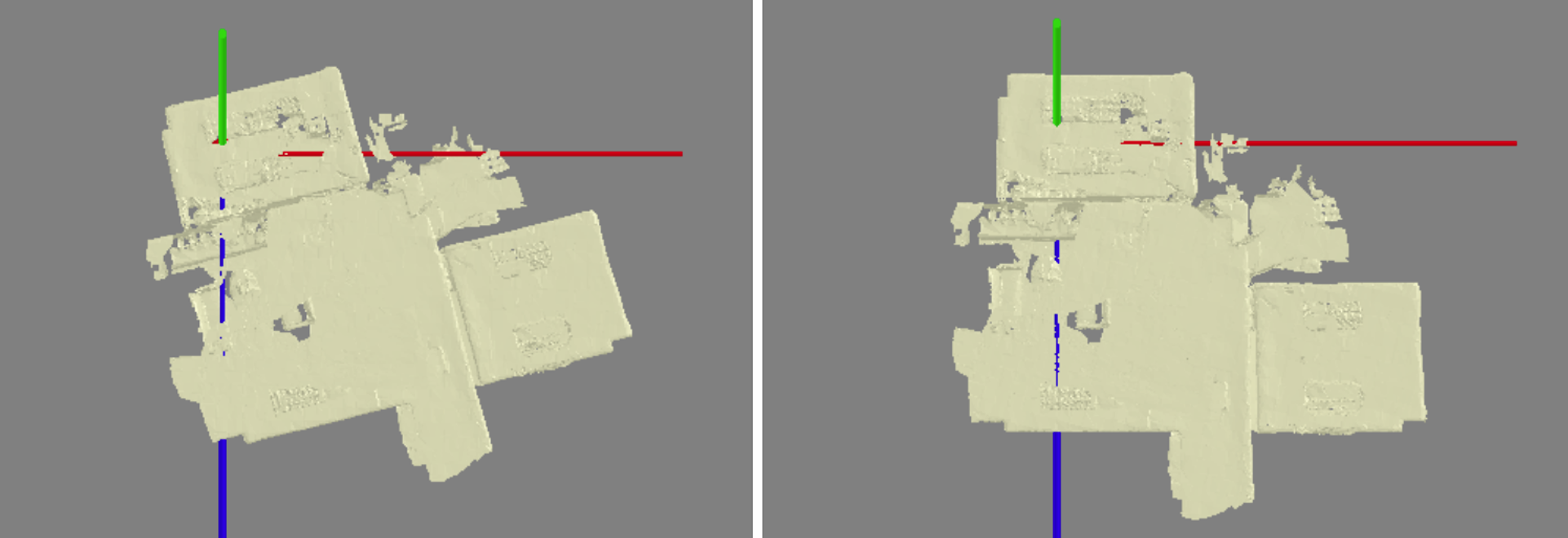}
  \centering
  \vspace{-0.1cm}
  \caption{The mesh model of building B1 (left) that is not aligned initially and the same after wall alignment (right).}
  \label{fig:Orientmesh-Experiment1}
\end{figure}
\vspace{-0.5cm}

Once the primary wall directions are computed, we pick the cluster with the largest number of triangles, take its direction (cluster center), project that direction onto the $x-z$ plane, and call it $\theta_{wall}$. We rotate the mesh by the angle between $\theta_{wall}$ and the $x$ axis. Now the primary walls will be pointing along the $x$ axis. Figure~\ref{fig:Orientmesh-Experiment1}-left shows a building that is not aligned with the axes. Figure~\ref{fig:Orientmesh-Experiment1}-right shows the same building after wall rotation. Adding in the x axis (in red) and z axis (in blue), we see that the walls are now well-aligned with the axes.

\vspace{-0.45cm}
\subsubsection{Dividing a mesh into separate levels}
\begin{figure}[!ht]
  \includegraphics[width=0.8\columnwidth]{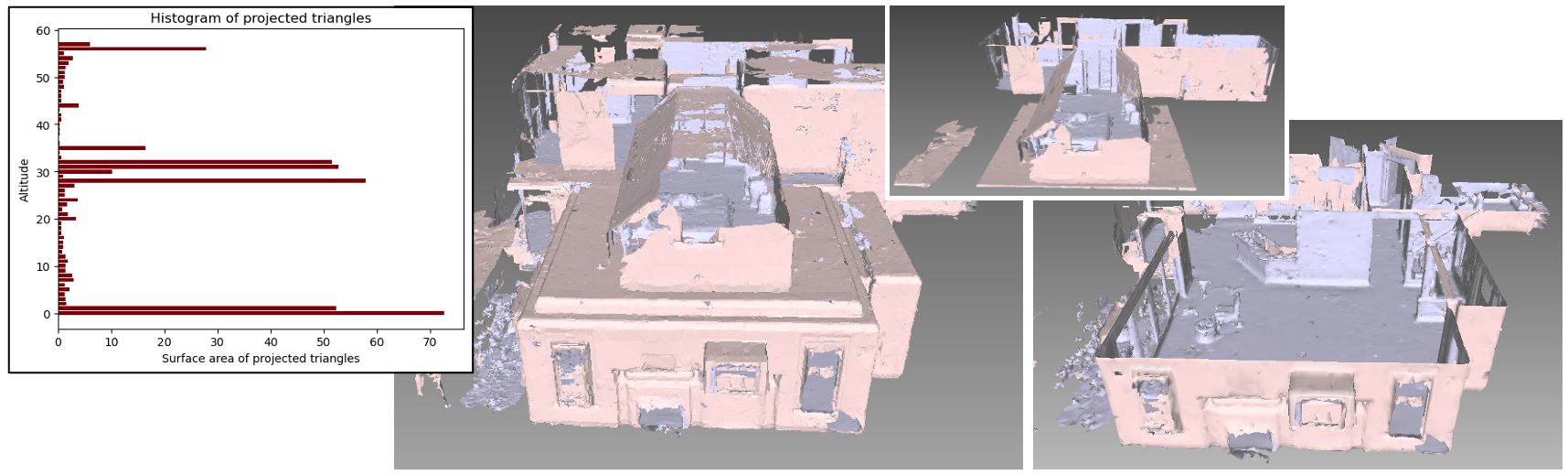}
  \vspace{-0.2cm}
  \centering
  \caption{Left: Using the triangles that face nearly straight up or straight down, plot a histogram of the centroid altitudes, weighted by triangle surface area. Histogram spikes are likely altitudes for floors and ceilings. Middle: A two-story building. Right: the two stories of that building.}
  \label{fig:histogram-horizontal}
\end{figure}

The HoloLens can digitize multi-story buildings. Given a multi-story model, we can compute a floor plan for each story. The process is similar to the one used in~\ref{sec:findheight} to find the height of the floor. First, our system computes a histogram, as shown in figure~\ref{fig:histogram-horizontal} and segments the building into multiple levels, as shown in figure~\ref{fig:histogram-horizontal}-middle and right.

\vspace{-0.35cm}
% \vspace{-0.25cm}

\subsection{Floor plan computation}
% \vspace{-0.1cm}
Our floor plan computation depends on the type of floor plan desired and whether the mesh is oriented with respect to the global axes. If we desire a pen-and-ink style floor plan and the mesh is oriented, we can simply pass the mesh $M$ to the \textit{ComputeAndSuperimposeSlices}() function, as in line 14 of Algorithm~\ref{alg:FPC}.

\vspace{-0.25cm}
\begin{algorithm}
\footnotesize{
\caption{Compute Floor Plan}
\begin{algorithmic}[1]
    \State $\Phi \gets \emptyset$
    \ForEach {$\Pi_i \in \Pi $}
        \State $\Delta_i \gets$ \textit{GetTrianglesFacingThisWay}($\mathcal{M}, Pi_i$)
        \State $\mathcal{C}_i \gets$ \textit{GetCentroids}($\Delta_i$)
        \State ${CL}_i \gets$ \textit{ModifiedDBSCAN}(${\mathcal{C}_i},l,w,h$)
         \State ${WS} \gets \emptyset$
         \ForEach {${CL}_{ij} \in {CL}_i$}
            \If{${CL}_{ij}$ \textit{extends floor to ceiling}}
              \State ${WS} \gets {WS} + {CL}_{ij}$
            \EndIf
         \EndFor
        \ForEach {${WS}_k \in {WS}$}
            \State $\mathcal{R} \gets$ \textit{ComputeRectangles}(${WS}_k$)
        \EndFor
        \State ${\Phi}_i \gets {\Phi}_i + \mathcal{R}$
    \EndFor
    \State ${M'} \gets$ \textit{AssembleWalls}$(\Phi)$
    \State $\mathcal{FP} \gets$\textit{ComputeAndSuperimposeSlices}($M'$)
\Return{$\mathcal{FP}$}
\end{algorithmic}
}
\label{alg:FPC}

\end{algorithm}

\vspace{-1.1cm}

\begin{figure}[!htb]
  \includegraphics[width=0.6\columnwidth]{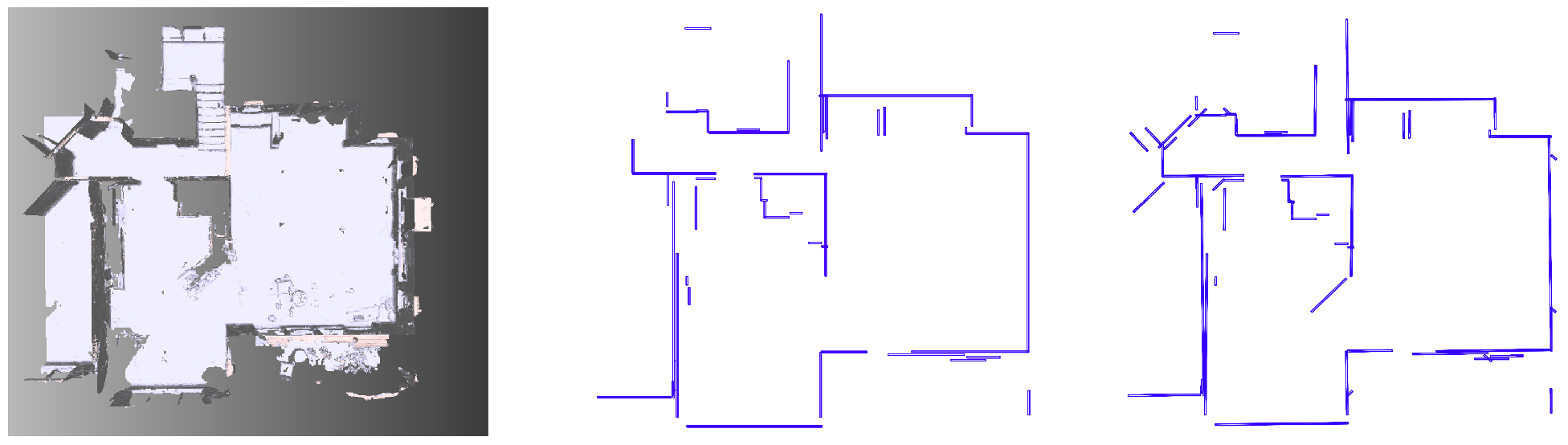}
  \vspace{-0.3cm}
  \centering
  \caption{Left: 3D model. Middle: floor plan obtained via spherical k-means with k=6. Right: floor plan from a modified DBSCAN that supports more wall orientations.}
  \label{fig:KMeansAgainstDBSCAN}
\end{figure}
\vspace{-0.5cm}

However, if the mesh is not properly oriented, we align it with the global axes before computing the floor plan. If a drafting-style floor plan is desired, we utilize lines 2-13 of Algorithm~\ref{alg:FPC} to compute flat walls. Then we slice the mesh to derive the floor plan using line 14 of the algorithm.

% Once the mesh is oriented with respect to the global axes, our floor plan computation is either a one-step or two-step process depending on the desired floor plan. Algorithm~\ref{alg:FPC} describes the main steps of our floor plan computation. To generate a drafting-style floor plan, we first compute flat walls (lines 2-13 in Algorithm~\ref{alg:FPC}) and then the mesh is sliced to generate the floor plan (line 14). To generate a pen-and-ink floor plan, we can skip the first step for computing flat walls; the oriented mesh $M$ is directly passed to \textit{ComputeAndSuperimposeSlices}() function as shown in line 14 of Algorithm~\ref{alg:FPC}.

\vspace{-0.25cm}
\subsubsection{Computing flat walls}
To generate a drafting-style floor plan, we compute flat walls and separate them from other building contents using these steps:

\textbf{Wall directions} Once the floor and ceiling are removed, the remaining walls consist of triangles having surface normals parallel to the $xz$ plane assuming $-y$ is the gravity direction. We use these surface normals to divide the triangles into k sets based on their facing directions $\Pi_i$, which we call \textit{wall directions}, denoted as $\Pi$. For our initial experiment, we consider $k = 4$ for the principal directions (North, East, South, and West). With these parameters, we are able to achieve precise floor plans in buildings where walls are along principal directions. However, this approach fails when the walls of the building face in arbitrary directions, as in figure~\ref{fig:KMeansAgainstDBSCAN}. In such cases, we use spherical k-means to find all the wall directions and then pass them to our DBSCAN algorithm.  

% If the building does not have walls other than the four principle direction, the user can provide these direction and apply DBSCAN. In case when the direction of the walls in the building are at arbitrary direction, then we search for all the direction by running K-means and 

\textbf{DBSCAN} For each wall direction, we perform a modified DBSCAN algorithm: We compute the centroid $\mathcal{C}$ of each triangle $\Delta_i$. For each centroid point $\mathcal{C}_i$ during DBSCAN, we count the number of other centroid points that are near enough to be considered neighbors. However, instead of looking for neighbors in a sphere around each point
%remove \textcolor{red} and paranthesis to change the color
as for traditional DBSCAN in 3D, we look for neighbors in a rectangular block of length $l$, width $w$ and height $h$ centered on the point. This block is tall enough to reach from floor to ceiling in the y direction, a little less wide than a door in the direction parallel to the proposed wall (e.g., 1.5 feet), and a few inches in the wall direction (to allow for walls that deviate slightly from being perfectly flat). 
%remove \textcolor{red} and paranthesis to change the color
Relying on the National Building Code, the wall's minimum height is set at 8 feet, with a thickness of 8 inches. After DBSCAN, the mesh triangles are grouped into wall segments $WS$.

\textbf{Filtering} We discard wall segments that are not good candidates, such as walls that are too small, that aren't near the floor, or that aren't near the ceiling.

\textbf{Plane fitting} For each wall segment, we find a plane that has the same facing direction as the wall direction and that is a good fit to the triangle centroids in that wall segment. Given that the points are tightly collected in this direction, simply having the plane go through any centroid works surprisingly well. However, it is also possible to choose a point more carefully, such as by finding a point at a median position in the wall direction.

\textbf{Rectangle construction} For each remaining wall segment, we construct rectangles $\mathcal{R}$ that lie in the fitted plane and are as wide as the wall segment triangles in width and as tall as the wall segment triangles in height.

\textbf{Mesh replacement} For floor plan construction, we discard the original mesh triangles and replace them with the new planar wall rectangles to serve as a de-cluttered mesh. If the subsequent steps use libraries that expect a triangle mesh, we use two adjacent right triangles in place of each rectangle.

\vspace{-0.55cm}
\begin{figure}[!htb]
  \includegraphics[width=0.4\columnwidth]{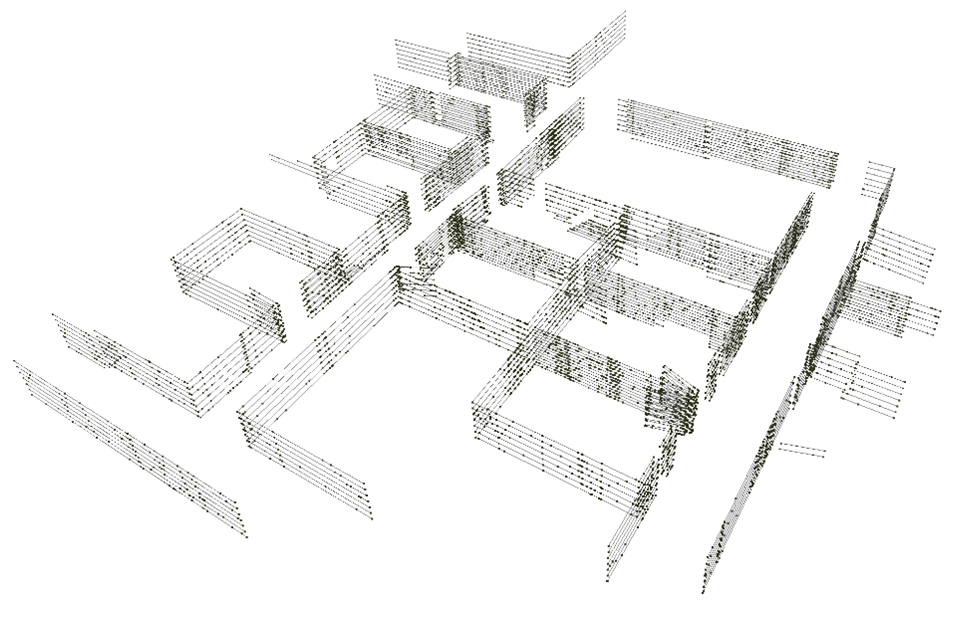}
  \vspace{-0.15cm}
  \centering
  \caption{A building mesh sliced at multiple altitudes.}
  \label{fig:slices}
\end{figure}
\vspace{-1.0cm}

\subsubsection{Slicing the mesh to produce a floor plan}
To produce a pen-and-ink style floor plan, we take our mesh after it has been rotated to have a level floor and to have walls aligned with the primary axes, and we slice that mesh at multiple altitudes. As described above, our histogram of $y$ values allows us to identify the height of the floor or floors and the height of the ceiling or ceilings. We pick a floor height, such as the highest floor, call it $y_{floor}$. We pick a ceiling height, such as the lowest ceiling, call it $y_{ceiling}$. Then we choose a series of $y$ values between $y_{floor}$  and $y_{ceiling}$, such as an even spacing of $y$ values:
\vspace{-0.3cm}
\begin{equation}
    y_i = y_{floor} + ( y_{ceiling} - y_{floor}) * (\frac{i}{n})
\end{equation}
for each $i$ such that $0 \leq i \leq n$. For each $y_i$, we compute the intersection of the mesh with the plane $y = y_i$. We end up with a stack of slices (see Figure~\ref{fig:slices}).

\vspace{-0.25cm}
\begin{figure}[!htb]
  \includegraphics[width=0.6\columnwidth]{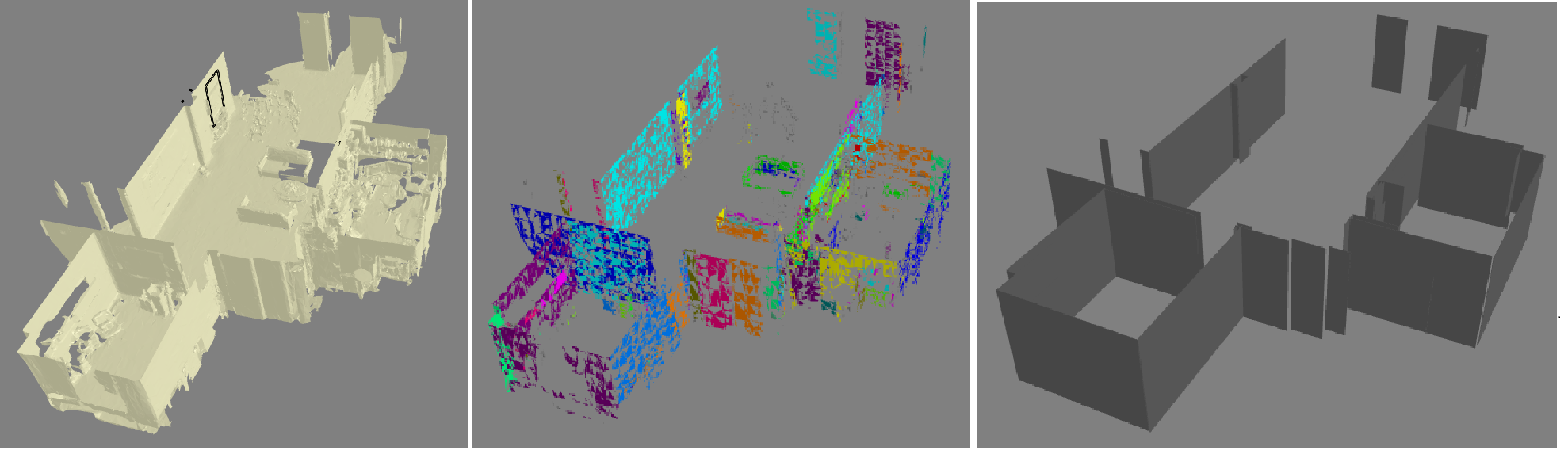}
  \centering
  \vspace{-0.15cm}
  \caption{Left: An oriented mesh from part of a building. Center: The results of DBSCAN. Each color represents a different wall segment. Right: The computed flat walls.}
  \label{fig:example1-mesh}
\end{figure}
\vspace{-0.75cm}

We can use the same method to produce a drafting-style floor plan. In this case, we begin with the flat wall model instead of the full mesh. This model has fewer details to capture by slicing at multiple altitudes, so we may choose to slice at a single intermediate altitude.

\vspace{-0.6cm}
\subsubsection{Drawing the floor plan}
For either style of floor plan, we can project the slices to a plane by ignoring the $y$ coordinates of the resulting line segments and plotting the resulting ($x$, $z$) coordinates as a two-dimensional image. We have also found it informative and aesthetically pleasing to draw the lines of each slice in a partially-transparent color, so that features that occur at multiple altitudes appear darker than features that occur only at a single altitude.

\begin{figure}[!ht]
  \centering
  \includegraphics[width=0.17\columnwidth, angle=90]{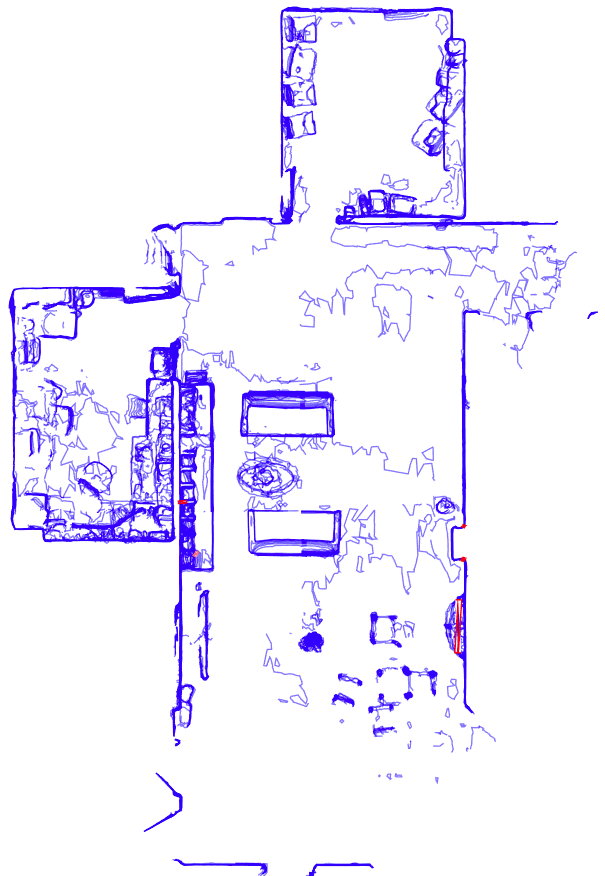}
  \hspace{0.25in}
    \includegraphics[width=0.17\columnwidth, angle=90] {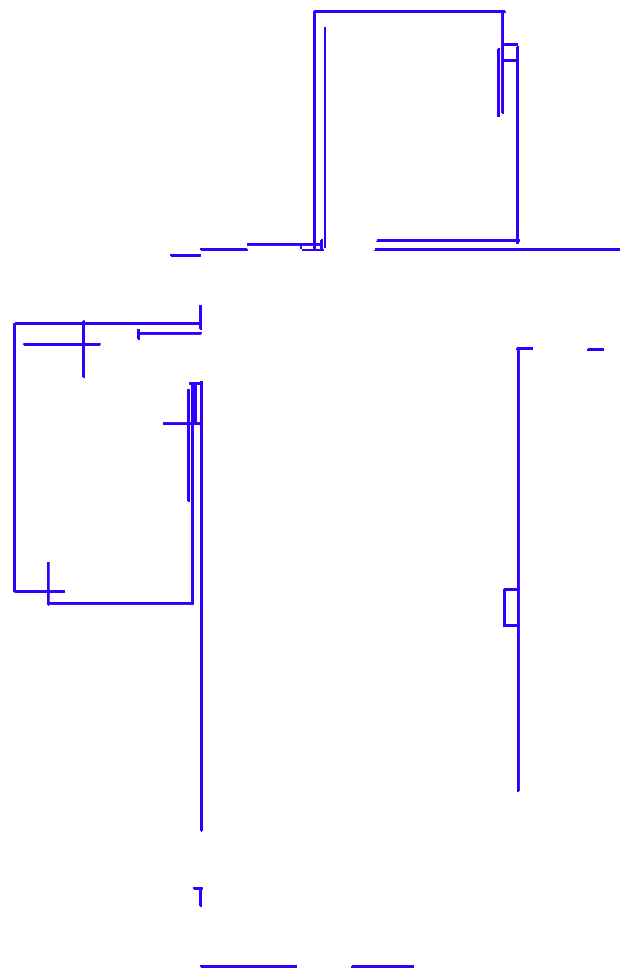}
    \vspace{-0.3cm}
  \caption{Pen-and-ink floor plan (left) and drafting floor plan (right) from Fig.~\ref{fig:example1-mesh}'s model.}
  \label{fig:pen-n-ink}
\end{figure}

As an example, Figure~\ref{fig:example1-mesh}-left shows a mesh gathered from a commercial building. Figure~\ref{fig:example1-mesh}-center shows the result of our DBSCAN on that data. Figure~\ref{fig:example1-mesh}-right shows the flat walls that result after mesh replacement. Figure~\ref{fig:pen-n-ink}-right shows the drafting-style floor plan that results from slicing the flat walls. Figure~\ref{fig:pen-n-ink}-left shows the pen-and-ink floor plan made by slicing the oriented mesh at multiple altitudes.
\vspace{-0.7cm}

\vspace{-1.2cm}
\subsubsection{Drawing synthetic objects}
Because our data comes from an AR headset, we can add synthetic objects to mark the positions of objects in a room, such as sensors and windows. We can display these objects in our 3D models and floor plans. For the steps of mesh processing described above, we note the geometric transformations applied to the mesh and apply the same transformations to the synthetic objects, which then appear in the correct places in the 3D views and in the floor plans. For example, the black objects in Figure~\ref{fig:example1-mesh}-left represent the objects placed by the user to show the positions of sensors and windows. Likewise, the red objects in the floor plan of Figure~\ref{fig:pen-n-ink}-left are those same objects, projected onto the same plane as the mesh slices.

% \vspace{-0.65cm}
% \begin{figure}[!ht]
%   \centering
%   \includegraphics[width=0.2\columnwidth, angle=90]{figures/pen-n-ink-completehouse.png}
%   \hspace{0.25in}
%     \includegraphics[width=0.2\columnwidth, angle=90] {figures/draftingStyleFloorPlan_Example1.png}
%     \vspace{-0.3cm}
%   \caption{Pen-and-ink floor plan (left) and drafting floor plan (right) obtained from the model of Fig.~\ref{fig:example1-mesh}}.
%   \label{fig:pen-n-ink}
% \end{figure}
% \vspace{-1.0cm}

%% file: 4-Experiments.tex
%=======================================
\section{Experiments}~\label{sec:exp}
%=======================================
% \vspace{-0.1cm}
We evaluated our approach to capturing 3D scans using AR headsets. We compared the floor plan dimensions with actual building dimensions and provided intermediate results: floor plans and 3D models. We also calculated the time taken for our algorithm steps. To demonstrate the approach robustness, we evaluated using multiple building types, including commercial buildings B1 and B3 and a residential building B3. Additionally, we validated our floor plan generation on the Matterport2D dataset~\cite{ramakrishnan2021hm3d}.

\vspace{-0.55cm}
\begin{figure}[!h]
  \includegraphics[width=0.65\columnwidth]{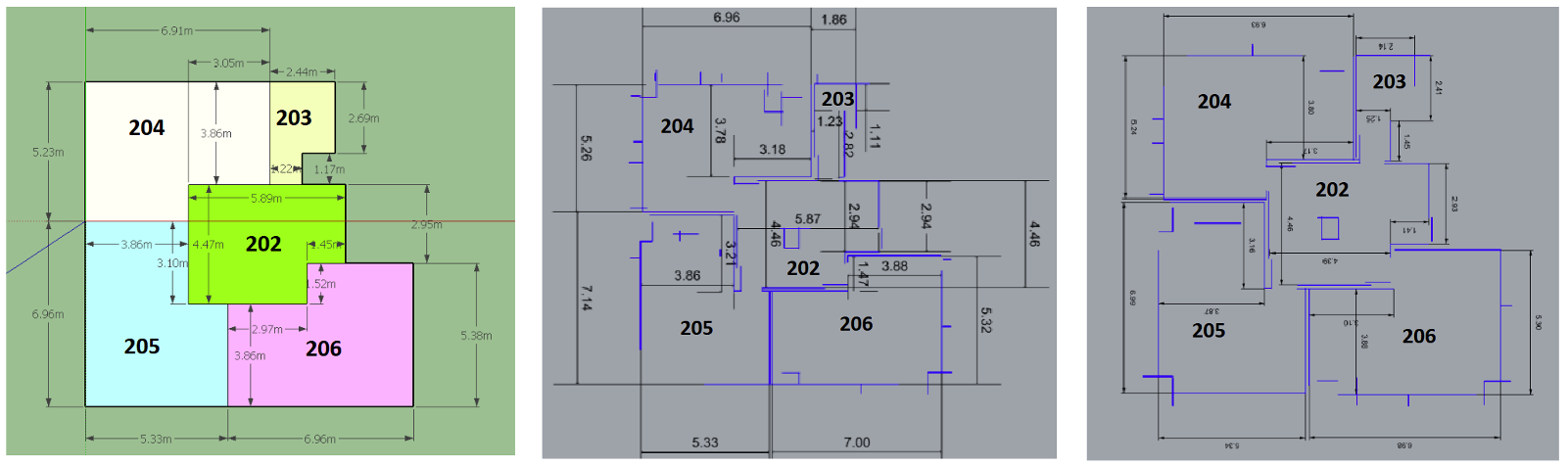}
  \centering
 \vspace{-0.25cm}
  \caption{ (a): Building measurement. (b): Scan S1 of the building. (c): Scan S2. }
  \label{fig:erroranalysis}
\end{figure}
\vspace{-0.5cm}

\textbf{Scanned Data Analysis}
We evaluated the precision of floor plan generation by comparing the actual dimensions of the rooms with the computed floor plan. Figure~\ref{fig:erroranalysis}(a) illustrates the measurement of the building, which was scanned twice with our AR. We call these scans S1 and S2 (see figure~\ref{fig:erroranalysis}(b) and figure~\ref{fig:erroranalysis}(c)) 
%remove \textcolor{red} and paranthesis to change the color
For each scan, floor plans are computed and the dimensions are computed using geometric modeling software, Rhino~\cite{rhino3d}. We then compared the computed dimensions with the actual room dimensions as shown in Table~\ref{tab:dimensionanalysis}. 
 %remove \textcolor{red} and paranthesis to change the color
These results show the applicability of our approach to many building types.

% \vspace{-0.6cm}
% \vspace{-0.2cm}
% \begin{figure*}[!ht]
%   \includegraphics[width=0.9\linewidth]{cgi2018_latex/figures/ComparisonForErrorAnalysis1.png}
%   \vspace{-0.3cm}
%   \centering
%   \caption{(a) shows the measurement of the building, while (b) and (c) depict the two different scans of the building, Scan 1 and Scan 2, respectively. }
%   \label{fig:erroranalysis}
% \end{figure*}
% \vspace{-0.4cm}

\vspace{-0.35cm}
\begin{table}[!ht]
\centering
\footnotesize{
\caption{Scan Accuracy: ${S1}{area}$ and ${S2}{area}$ represent the estimated area of each room for two scans. ${S1}_{err}$ and ${S2}_{err}$ are percentage error compared to actual dimensions.}
\vspace{-0.1cm}
\begin{tabular}{c | rrrrr} % creating five columns
\hline\hline %inserting double-line
Room & Area & ${S1}_{area}$ & ${S1}_{err} \%$ & ${S2}_{area}$ & ${S2}_{err} \%$\\ \hline
202 & 24.1243 & 24.037 & 0.4 & 32.7107 & 1.7\\ 
203 & 7.991 &5.5332 & 30.8 & 6.9699 & 12.8\\ 
204 & 31.9608 & 31.9032 & 0.2 & 31.7484 & 0.7\\ 
205 & 32.5398 & 33.3375 & -2.5 & 32.6814 & -0.4\\ 
206 & 32.9304 & 32.6536 & 0.8 & 32.592 & 1.0\\ 
\hline % inserts single-line
\end{tabular}
\label{tab:dimensionanalysis}
}
\end{table}
\vspace{-1.15cm}

\begin{figure}[!htb]
  \includegraphics[width=0.75\columnwidth]{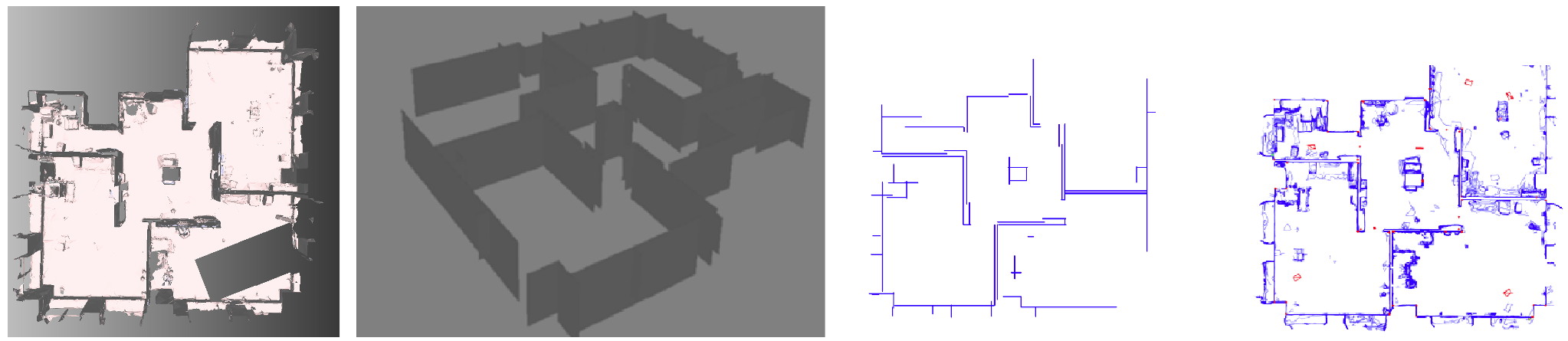}
 \vspace{-0.2cm}
  \centering
  \caption{First: 3D model of building B3. Second: computed flat walls. Third: drafting style floor plan. Fourth: pen-and-ink style floor plan.}
  \label{fig:ornl_floorplan}
\end{figure}
\vspace{-0.45cm}
%% commenting below to reduce text length - Ritesh

% % \vspace{-1.2cm}
% \begin{figure}[!ht]
%   \includegraphics[width=0.45\columnwidth]{figures/ResidentialMeshWithAndWithoutAlignment.png}
%   \vspace{-0.3cm}
%   \centering
%   \caption{Mesh model of building B2: Before alignment (left) and after alignment (right).}
%   \label{fig:Orientmesh-Experiment2}
% \end{figure}

%%%%%%%%%%%%%%%%%%%%%%%%%%%%%%%%%%%%%

Our method can compute a floor plan even from relatively incomplete mesh data. With a higher quality HoloLens scan, the resulting floor plan is more precise. Figure~\ref{fig:ornl_floorplan} displays S1 results: both types of floor plan and the 3D model.

\textbf{Orienting floor and walls} We must orient the mesh properly. Spherical k-means is compute intensive so we optimize it to get good performance. In Figure~\ref{fig:Orientmesh-Experiment1}, we see the mesh of B1 before and after alignment, which took 12.4 seconds, of which 10.6 were spent aligning walls using spherical k-means. 
%% commenting below to reduce text length - Ritesh
% Similarly, Figure~\ref{fig:Orientmesh-Experiment2} shows the mesh from building B2 before and after alignment.
%%%%%%%%%%%%%%%%%%%%%%%%%%%%%%%%%%%%%

\begin{figure}[!ht]
  \includegraphics[width=0.7\columnwidth]{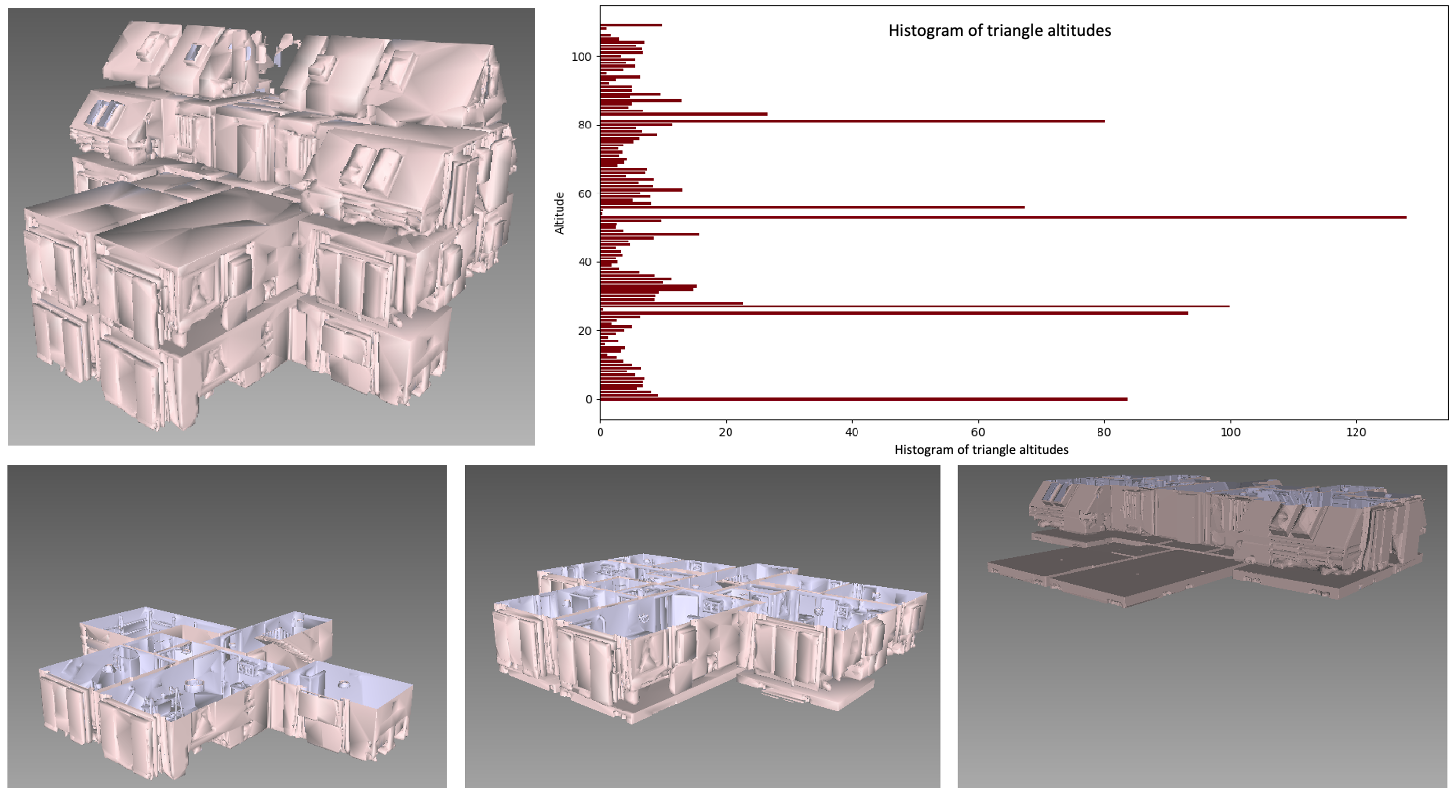}
  \vspace{-0.3cm}
  \centering
  \caption{A three-story model from Matterport3D (top-left) and its triangle altitude histogram (top-right). The bottom figure shows the building sliced into levels.}
  \label{fig:Level-Slicing-Experiment2}
\end{figure}

% \vspace{-0.5cm}
\textbf{Partitioning into stories} We can detect a multi-story building and divide it into stories with an additional step. The algorithm projects triangles onto the positive y-axis and creates a histogram showing horizontal peaks. By analyzing the peaks in the histogram, we can determine the number of stories. Figures~\ref{fig:histogram-horizontal} and~\ref{fig:Level-Slicing-Experiment2} show a 2-story residential building and a multi-story model from the Matterport3D dataset~\cite{ramakrishnan2021hm3d} that were partitioned into stories.

\vspace{-0.5cm}
\vspace{-0.5cm}

\begin{figure}[!ht]
  \includegraphics[width=0.75\columnwidth]{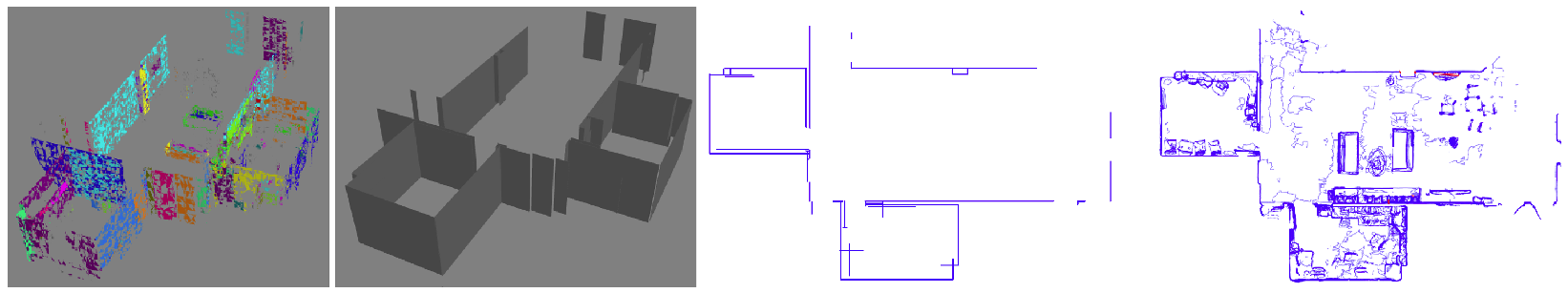}
  \vspace{-0.25cm}
  \centering
  \caption{First: DBSCAN clustering results for building B2. Second: plane-fitted flat walls. Third: drafting-style floor plan. Fourth: pen-and-ink style floor plan.}
  \label{fig:PlanarWalls-Experiment1}
\end{figure}

\begin{figure}[!htb]
  \includegraphics[width=0.75\columnwidth]{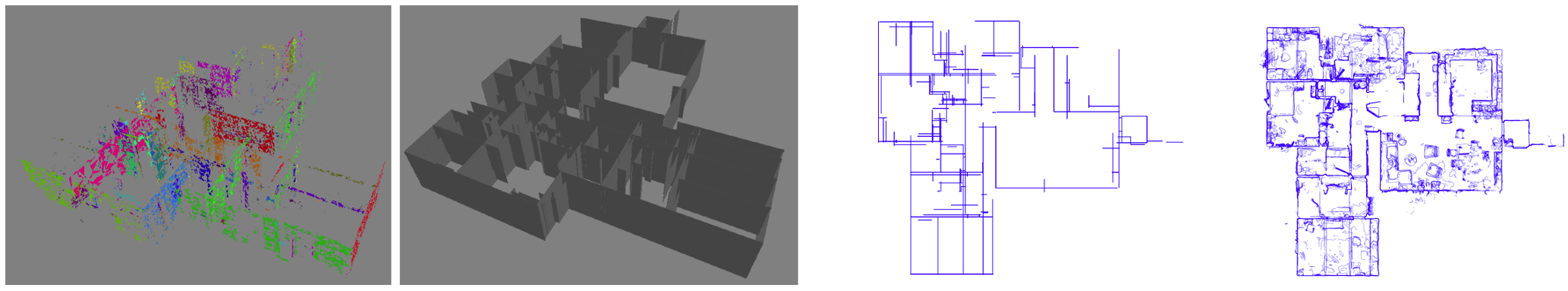}
  \centering
  \vspace{-0.15cm}
  \caption{First: DBSCAN clustering results for B2. Second: plane-fitted flat walls. Third: drafting-style floor plan. Fourth: (c) pen-and-ink style floor plan.}
  \label{fig:PlanarWalls-Experiment2}
\end{figure}

\begin{figure}[!ht]
  \includegraphics[width=0.75\columnwidth]{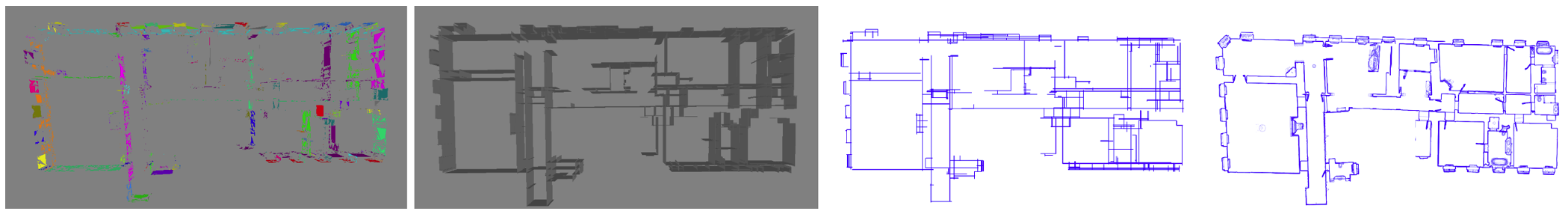}
  \centering
  \vspace{-0.25cm}
  \caption{First: DBSCAN clustering results for a model from Matterport3D. Second: plane-fitted flat walls. Third: drafting-style floor plan. Fourth: pen-and-ink floor plan.}
  \label{fig:PlanarWalls-Experiment3}
\end{figure}

\textbf{Finding planar walls} To generate a drafting-style floor plan, we eliminate details and identify planar walls. The modified DBSCAN algorithm is the most time-consuming step. In the model of Figure~\ref{fig:PlanarWalls-Experiment1}, with 79,931 vertices and 134,235 faces, it took 27.4 seconds to prepare the data and run DBSCAN and an additional 3.79 seconds to construct flat walls from the generated clusters. For the residential building of Figure~\ref{fig:PlanarWalls-Experiment2}, with 173,941 vertices and 285,840 faces, it took 76 seconds to prepare and run DBSCAN and 23.36 seconds to compute flat walls. The results for a Matterport3D model appear in Figure~\ref{fig:PlanarWalls-Experiment3}.

\vspace{-0.55cm}
\vspace{-0.75cm}

\textbf{Generating the floor plan} The final step of our floor plan generation is to slice the mesh at different heights and superimpose the slices. Figures~\ref{fig:PlanarWalls-Experiment1}, ~\ref{fig:PlanarWalls-Experiment2}, and~\ref{fig:PlanarWalls-Experiment3} show floor plans generated using our approach.

We conducted experiments to evaluate the effects of changing graphical settings when rendering pen-and-ink floor plans. Each setting consists of a different combination of line segment opacity and slice count. We found that an opacity setting of 0.5 produced a floor plan that met our expectations. We also found that a floor plan with 100 slices provided a good balance between level of detail and clutter reduction. Optimal numbers will depend on use case.

\vspace{-0.5cm}

\vspace{-1.4cm}

\vspace{-1.1cm}
\begin{table}[!ht]
\centering
\footnotesize{
\caption{Computation (Seconds): $T_o$, $T_p$, $T_{sd}$ are times for orienting the mesh, finding planar walls, and slicing/drawing.} %title of the table
\begin{tabular}{c | rrrr} % creating three columns
\hline\hline %inserting double-line
Building & Type & $T_o$ & $T_p$ &  $T_{sd}$\\ \hline
B1 & Commercial & 12.1 & 31.2  &  0.5\\
B2 & Residential &  24.08 & 81.5 & 0.9  \\ 
B3 & Commercial & 4.63 & 15.17  &  0.51\\ 
\hline % inserts single-line
\end{tabular}
\label{tab:analysis}
}
\end{table}
\vspace{-0.5cm}

Table~\ref{tab:analysis} lists elapsed time (in seconds) for orienting the mesh $T_o$, finding flat walls $T_p$ and slicing and drawing $T_{sd}$ for buildings B1, B2 and B3. 

% Add blown-up version of the furniture

% Table~\ref{tab:floorplanDrawingAnalysis} shows an analysis for choosing the estimate on the number of slices required for drawing floor plan. It can be seen from Table~\ref{tab:floorplanDrawingAnalysis} for scene 1, one-third of the requested slices ends up well formed and non-empty.

% \vspace{-0.25cm}
% \begin{figure}[!ht]
%   \includegraphics[width=0.3\columnwidth, angle=90]{figures/draftingStyleFloorPlan_Example1.png}
%   \includegraphics[width=0.3\columnwidth, angle=90]{figures/pen-n-ink-completehouse.png}
%   \centering
%   \caption{Two styles of floor plan obtained for B1 building: Drafting-style (left) and pen-and-ink style (right).}
%   \label{fig:floorplans-Experiment1}
% \end{figure}
% \vspace{-0.25cm}

% \begin{figure}[!ht]
%   \includegraphics[width=0.70\columnwidth]{figures/Residential_Floorplans.png}
%   \centering
%   \caption{Two styles of floor plan obtained for B2: Drafting-style floor plan (left) and  pen-and-ink style floor plan (right).}
%   \label{fig:floorplans-Experiment2}
% \end{figure}
% \begin{figure}[!ht]
%   \includegraphics[width=0.9\columnwidth]{figures/floorplan_experiment3b.png}
%   \centering
%   \vspace{-0.2cm}
%   \caption{Two styles of floor plan obtained for the Matterport3D Dataset: Drafting-style floor plan (left) and pen-and-Ink style floor plan (right).}
%   \label{fig:floorplans-Experiment3}
% \end{figure}
% \vspace{-0.25cm}

%% file: 5-conclusion.tex
\section{Conclusion \& Future work}
\vspace{-0.25cm}
In summary, our new approach for generating floor plans from triangle mesh data collected by augmented reality (AR) headsets produces two styles: a detailed pen-and-ink style and a simplified drafting style. Our algorithms align the mesh data with primary coordinate axes to produce tidy floor plans with vertical and horizontal walls, while also allowing for the removal of ceilings and floors and the separation of multi-story buildings into individual stories. Our approach integrates with AR, supporting the addition of synthetic objects to physical geometry and providing a detailed 3D model and floor plan.
% Our approach for generating floor plans from triangle mesh data collected by augmented reality headsets has yielded two distinct styles of floor plans: a detailed pen-and-ink style and a simplified drafting-style that focuses on planar walls. By aligning the mesh data with primary coordinate axes early in the process, we are able to produce tidy-looking floor plans with vertical and horizontal walls. We have also developed methods to remove ceilings and floors and separate multi-story buildings into individual stories. Our approach integrates with augmented reality, allowing for the addition of synthetic objects to physical geometry, which can be viewed in the 3D models and floor plans we generate.

Potential applications include navigation, interior design, furniture placement, facility management, building construction, and HVAC design. Moving forward, we plan to enable support for sloping ceilings, automate wall and door detection, and integrate with other tools such as energy simulators. Finally, we plan to compare our approach with existing state-of-the-art methods in terms of accuracy and computational time. We also plan to explore the applicability of block-based DBScan for 3D reconstruction from incomplete scans. Our approach has the potential to revolutionize the way we generate and visualize floor plans.

% We are continuously improving our floorplan generation algorithms and have achieved promising results, generating pen-and-ink floor plans for multi-room models in under 15 seconds, starting with a triangle mesh straight from a HoloLens. Our floor plans have potential applications in various fields, including navigation, interior design, furniture placement, facilities management, building construction, and HVAC design.

% We plan to expand our work by improving performance, providing better support for sloping ceilings, automatically detecting walls and doors, and integrating with other tools such as building energy simulators. Overall, our approach contributes to the field of computer-aided design and has the potential to revolutionize the way we generate and visualize floor plans.

% \section*{Acknowledgements}

% To be added at publication time.

%DO NOT FORGET: Acknowledgements:

%% file: samplepaper.bbl
\begin{thebibliography}{10}
\providecommand{\url}[1]{\texttt{#1}}
\providecommand{\urlprefix}{URL }
\providecommand{\doi}[1]{https://doi.org/#1}

\bibitem{Adan2010}
Adan, A., Huber, D.: 3d reconstruction of interior wall surfaces under occlusion and clutter. In: 2011 International Conference on 3D Imaging, Modeling, Processing, Visualization and Transmission. pp. 275--281 (2011). \doi{10.1109/3DIMPVT.2011.42}

\bibitem{Arikan2013}
Arikan, M., Schw\"{a}rzler, M., Fl\"{o}ry, S., Wimmer, M., Maierhofer, S.: O-snap: Optimization-based snapping for modeling architecture. ACM Trans. Graph.  \textbf{32}(1) (feb 2013). \doi{10.1145/2421636.2421642}

\bibitem{Budroni2010}
Budroni, A., Boehm, J.: Automated 3d reconstruction of interiors from point clouds. International Journal of Architectural Computing  \textbf{8}(1),  55--73 (2010). \doi{10.1260/1478-0771.8.1.55}

\bibitem{Cabral2014}
Cabral, R.S., Furukawa, Y.: Piecewise planar and compact floorplan reconstruction from images. 2014 IEEE Conference on Computer Vision and Pattern Recognition pp. 628--635 (2014)

\bibitem{cai2022}
Cai, R., Li, H., Xie, J., Jin, X.: Accurate floorplan reconstruction using geometric priors. Computers \& Graphics  \textbf{102},  360\--369 (2022). \doi{10.1016/j.cag.2021.10.011}

\bibitem{Chen_2019}
Chen, J., Liu, C., Wu, J., Furukawa, Y.: Floor-sp: Inverse cad for floorplans by sequential room-wise shortest path. In: The IEEE International Conference on Computer Vision (ICCV) (2019)

\bibitem{Chen2022}
Chen, N., Lu, Z., Yu, X., Yang, L., Xu, P., Fan, Y.: Augmented reality-based home interaction layout and evaluation. In: Computer Graphics International Conference. pp. 395--406. Springer (2022)

\bibitem{dasgupta2016}
Dasgupta, S., Fang, K., Chen, K., Savarese, S.: Delay: Robust spatial layout estimation for cluttered indoor scenes. In: 2016 IEEE Conference on Computer Vision and Pattern Recognition (CVPR). pp. 616--624 (2016). \doi{10.1109/CVPR.2016.73}

\bibitem{Furukawa2009}
Furukawa, Y., Curless, B., Seitz, S.M., Szeliski, R.: Reconstructing building interiors from images. In: 2009 IEEE 12th International Conference on Computer Vision. pp. 80--87 (2009). \doi{10.1109/ICCV.2009.5459145}

\bibitem{Gao2014}
Gao, R., Zhao, M., Ye, T., Ye, F., Wang, Y., Bian, K., Wang, T., Li, X.: Jigsaw: Indoor floor plan reconstruction via mobile crowdsensing. In: Proceedings of the 20th Annual International Conference on Mobile Computing and Networking. p. 249–260. MobiCom '14, Association for Computing Machinery, New York, NY, USA (2014). \doi{10.1145/2639108.2639134}

\bibitem{Hsiao2019}
Hsiao, C.W., Sun, C., Sun, M., Chen, H.T.: Flat2layout: Flat representation for estimating layout of general room types. ArXiv  \textbf{abs/1905.12571} (2019)

\bibitem{Ikehata2015}
Ikehata, S., Yang, H., Furukawa, Y.: Structured indoor modeling. In: 2015 IEEE International Conference on Computer Vision (ICCV). pp. 1323--1331 (2015). \doi{10.1109/ICCV.2015.156}

\bibitem{ivan2020}
Kruzhilov, I., Romanov, M., Babichev, D., Konushin, A.: Double refinement network for room layout estimation. In: Palaiahnakote, S., Sanniti~di Baja, G., Wang, L., Yan, W.Q. (eds.) Pattern Recognition. pp. 557--568. Springer International Publishing, Cham (2020)

\bibitem{Lee2017}
Lee, C.Y., Badrinarayanan, V., Malisiewicz, T., Rabinovich, A.: Roomnet: End-to-end room layout estimation. 2017 IEEE International Conference on Computer Vision (ICCV) pp. 4875--4884 (2017)

\bibitem{Liu2018}
Liu, C., Wu, J., Furukawa, Y.: Floornet: A unified framework for floorplan reconstruction from 3d scans. In: ECCV (2018)

\bibitem{Liu2013}
Liu, H., Yang, Y.L., AlHalawani, S., Mitra, N.J.: Constraint-aware interior layout exploration for precast concrete-based buildings. Visual Computer (CGI Special Issue)  (2013)

\bibitem{rhino3d}
McNeel, R., et~al.: Rhinoceros 3d, version 6.0. Robert McNeel \& Associates, Seattle, WA  (2010)

\bibitem{microsoft_spatial_mapping}
Microsoft: Spatial mapping. \url{https://docs.microsoft.com/en-us/windows/mixed-reality/spatial-mapping} (2022)

\bibitem{Monszpart2015}
Monszpart, A., Mellado, N., Brostow, G.J., Mitra, N.J.: Rapter: Rebuilding man-made scenes with regular arrangements of planes. ACM Trans. Graph.  \textbf{34}(4) (jul 2015). \doi{10.1145/2766995}

\bibitem{mura2016}
Mura, C., Mattausch, O., Pajarola, R.: Piecewise-planar reconstruction of multi-room interiors with arbitrary wall arrangements. Computer Graphics Forum  \textbf{35}(7),  179--188 (2016). \doi{https://doi.org/10.1111/cgf.13015}

\bibitem{murali2017}
Murali, S., Speciale, P., Oswald, M.R., Pollefeys, M.: Indoor scan2bim: Building information models of house interiors. In: 2017 IEEE/RSJ International Conference on Intelligent Robots and Systems (IROS). pp. 6126--6133 (2017). \doi{10.1109/IROS.2017.8206513}

\bibitem{Okorn2010}
Okorn, B., Xiong, X., Akinci, B.: Toward automated modeling of floor plans. In: In Proceedings of the symposium on 3D data processing, visualization and transmission. vol.~2 (2010)

\bibitem{Pintore2014}
Pintore, G., Gobbetti, E.: Effective mobile mapping of multi-room indoor structures. The visual computer  \textbf{30}(6-8),  707--716 (2014)

\bibitem{Pintore2020}
Pintore, G., Mura, C., Ganovelli, F., Fuentes-Perez, L.J., Pajarola, R., Gobbetti, E.: {State-of-the-art in Automatic 3D Reconstruction of Structured Indoor Environments}. Computer Graphics Forum  (2020). \doi{10.1111/cgf.14021}

\bibitem{ramakrishnan2021hm3d}
Ramakrishnan, S.K., Gokaslan, A., Wijmans, E., Maksymets, O., Clegg, A., Turner, J.M., Undersander, E., Galuba, W., Westbury, A., Chang, A.X., Savva, M., Zhao, Y., Batra, D.: Habitat-matterport 3d dataset ({HM}3d): 1000 large-scale 3d environments for embodied {AI}. In: Thirty-fifth Conference on Neural Information Processing Systems Datasets and Benchmarks Track (Round 2) (2021), \url{https://openreview.net/forum?id=-v4OuqNs5P}

\bibitem{Turner2012}
Turner, E., Zakhor, A.: Watertight as-built architectural floor plans generated from laser range data. In: 2012 Second International Conference on 3D Imaging, Modeling, Processing, Visualization Transmission. pp. 316--323 (2012). \doi{10.1109/3DIMPVT.2012.80}

\bibitem{weinmann2021efficient}
Weinmann, M., Wursthorn, S., Weinmann, M., H{\"u}bner, P.: Efficient 3d mapping and modelling of indoor scenes with the microsoft hololens: A survey. PFG--Journal of Photogrammetry, Remote Sensing and Geoinformation Science  \textbf{89}(4),  319--333 (2021)

\bibitem{XIONG2013}
Xiong, X., Adan, A., Akinci, B., Huber, D.: Automatic creation of semantically rich 3d building models from laser scanner data. Automation in Construction  \textbf{31},  325--337 (2013). \doi{10.1016/j.autcon.2012.10.006}

\bibitem{zhang2013}
Zhang, J., Kan, C., Schwing, A.G., Urtasun, R.: Estimating the 3d layout of indoor scenes and its clutter from depth sensors. In: 2013 IEEE International Conference on Computer Vision. pp. 1273--1280 (2013). \doi{10.1109/ICCV.2013.161}

\bibitem{Zou2018}
Zou, C., Colburn, A., Shan, Q., Hoiem, D.: Layoutnet: Reconstructing the 3d room layout from a single rgb image. In: 2018 IEEE/CVF Conference on Computer Vision and Pattern Recognition (CVPR). pp. 2051--2059. IEEE Computer Society, Los Alamitos, CA, USA (jun 2018). \doi{10.1109/CVPR.2018.00219}

\end{thebibliography}
